\documentclass[twocolumn,epsf,prd]{revtex4}
\usepackage{graphicx}
\usepackage{dcolumn}
\usepackage{bm}
\begin{document}

\newcommand{\be}{\begin{equation}}
\newcommand{\ee}{\end{equation}}
\newcommand{\ba}{\begin{eqnarray}}
\newcommand{\ea}{\end{eqnarray}}

\title{Ultra-High Energy Heavy Nuclei Propagation in Extragalactic
Magnetic Fields}

\author{Gianfranco Bertone$^a$, Claudia Isola$^{a,b}$, Martin Lemoine$^a$,
G{\"u}nter Sigl$^a$}

\affiliation{$^a$ GReCO, Institut d'Astrophysique de Paris, C.N.R.S.,
98 bis boulevard Arago, F-75014 Paris, France\\
$^b$ Centre de Physique Th{\'e}orique,Ecole Polytechnique,
91128 Palaiseau Cedex, France\\}

\vspace{0.5truecm}
\begin{abstract}
We extend existing work on the propagation of ultra-high energy cosmic
rays in extragalactic magnetic fields to a possible component of heavy
nuclei, taking into account photodisintegration, pion production, and
creation of $e^{\pm}$ pairs. We focus on the influence of the magnetic
field on the spectrum and chemical composition of observed ultra-high
energy cosmic rays. We apply our simulations to the scenarios proposed
by Anchordoqui {\it et al.}, in which Iron nuclei are accelerated in
nearby starburst galaxies, and show that it is in marginal agreement
with the data. We also show that it is highly unlikely to detect He
nuclei from M87 at the highest energies observed $\sim3\,10^{20}\,$eV
as required for the scenario of Ahn {\it et al.} in which the highest
energy cosmic rays originate from M87 and are deflected in a Parker
spiral Galactic magnetic field.
\end{abstract}

\maketitle
\vspace{1truecm}

\section{Introduction}
The origin of ultra-high energy cosmic rays (UHECR) is one of the
major open questions in astro-particle physics.
Data from the Fly's Eye experiment \cite{Fly} suggest that the
chemical composition is dominated by heavy nuclei up to the ankle
($E\simeq10^{18.5}\,$eV) and then progressively by protons beyond,
while other data \cite{Hayashida} may suggest a mixed composition of
both protons and heavier nuclei.  The fact that present experiments do
not give a clear answer to the question of chemical composition of
primary particles motivates to test scenarios with a heavy component.

Nucleons cannot be confined in our galaxy at energies above the ankle;
together with the absence of a correlation between their arrival
directions and the galactic plane, this suggests that if nucleons are
primary particles they should have an extragalactic origin.  At the
same time, nucleons at energies above $\simeq4\times 10^{19}$ eV
interact with the photons of the Cosmic Microwave Background (CMB) by
photopion production; this would predict a break in the cosmic ray
flux, the so-called GZK cut-off~\cite{gzk}, and the sources of UHECR
above the GZK cut-off should be nearer than about $50\,$Mpc.  The GZK
cut-off has not been observed by the experiments such as Fly's
Eye~\cite{Fly}, Haverah Park~\cite{Haverah}, Yakutsk~\cite{Yakutsk},
and AGASA~\cite{AGASA}. However, currently there seems to be a
disagreement specifically between the AGASA ground array~\cite{AGASA}
which detected about 10 events above $10^{20}\,$eV, as opposed to
about 2 expected from the GZK cut-off, and the HiRes fluorescence
detector~\cite{hires} which seems consistent with a
cut-off~\cite{wb}. The resolution of this problem may have to await
the completion of the Pierre Auger project~\cite{auger} which will
combine the two existing complementary detection techniques.

In the acceleration scenario, UHECR can achieve these extreme high
energies by acceleration in shocked magnetized plasmas in powerful
astrophysical sources, such as hot spots of radio galaxies and active
galactic nuclei\cite{biermann}.

Attributing sources to the highest energy events is complicated by the
lack of observed counterparts~\cite{ssb,ES95}. A possible explanation
is the existence of large scale intervening magnetic fields with
intensities $B\sim0.1-1\,\mu$G~\cite{ES95}, which would provide
sufficient angular deflection even for high energies and could explain
the large scale isotropy of arrival directions observed by the AGASA
experiment~\cite{AGASA} as due to diffusion. In this framework, the
clusters of events seen by the AGASA and Yakutsk
experiments~\cite{AGASA,TT} are interpreted as due to focussing of the
highest energy cosmic rays in caustics of the extra-galactic magnetic
fields, as originally suggested in Ref.~\cite{LSB99} (see also
Ref.~\cite{HMR99} for nuclei propagating in the Galactic magnetic
field and Ref.~\cite{HMR02} for recent detailed analytical
studies). Indeed it has been realized recently that magnetic fields as
strong as $\simeq 1 \mu G$ in sheets and filaments of large scale
structures, such as our Local Supercluster, are compatible with
existing upper limits on Faraday rotation~\cite{vallee,ryu,blasi}.

Heavy nuclei as UHECR primaries are interesting in two ways in this
context : they can be accelerated more easily to high energies, as the
maximal acceleration energy a particle can achieve depends linearly on
its charge ${\rm Ze}$, and, in addition, the increased deflection
(also proportional to ${\rm Ze}$), could explain more easily the
absence of correlation between the arrival direction of the events and
the nearest powerful astrophysical objects. However, even in this case
there is a limit on the distance to the source because of
photodisintegration processes due to the interaction with infra-red
and CMB.

The study of the propagation of heavy nuclei in the absence of
magnetic deflection has been treated in some detail in the
literature. The pioneering work of Puget, Stecker and Bredekamp (PSB
in the following)~\cite{PSB} which included all energy loss
mechanisms, has been recently updated~\cite{Epele,salamon} to take
into account new empirical estimates of the infrared background
density of photons~\cite{Malkan} which are about one order of
magnitude lower than used by PSB.

In this paper we study the propagation of a distribution of heavy
nuclei in a stochastic magnetic field, including all relevant
energy loss processes. Our numerical simulations allow to treat in
a consistent way the interplay between magnetic deflection and
photodisintegration losses. We also keep track of the propagation
of all nucleon secondaries produced in photodisintegration events,
and propagate these secondaries in the magnetic field. These
effects had not been considered in previous studies of UHE nuclei
propagation. In particular, we focus here on the influence of the
magnetic field on the observable UHECR spectrum and its chemical
composition.  In contrast to the sky distribution, these
quantities are not significantly influenced by Galactic magnetic
fields which we therefore neglect. As will be seen in what
follows, a relatively strong magnetic field ($B\gtrsim
10^{-8}\,$G), {\it i.e.} such that UHECR of low energies diffuse,
modify by its presence the chemical composition and the energy
spectrum recorded at a given distance. This is due to the effect
of diffusion, which increases the local residence time
differentially with energy, as well as the effective length
traveled hence the photodisintegration probability. The interplay
between these effects is rather complex, and the output spectrum
and chemical composition thus depend on several parameters such as
the maximum injection energy, injection spectral index, linear
distance and initial chemical composition.  Due to the rather high
dimensionality of the parameter space, we will show results for
fixed values of the maximum injection energy $E_{\rm
max}=10^{22}\,$eV and spectral index ${\rm d}n/{\rm d}E\propto
E^{-2}$, at the expense of generality, and discuss how the
conclusions would be modified for other values of these
parameters.

The paper is organized as follows: in section II we describe the
propagation of UHE heavy nuclei, in section III we describe our
numerical simulation, in section IV we present our results, in section
V we apply our results to test the validity of some recent models, and
in section VI we conclude.

\section{Energy Loss Rates}
Heavy nuclei are attenuated basically by two processes:
photodisintegration on the diffuse photon backgrounds and creation of
$e^{\pm}$ pairs~\cite{PSB,Stecker,ES95}. For energies above
$10^{20}\,$eV, it is the CMB which mostly contributes to the
photodisintegration process, whereas at lower energies the infra-red
background provides the main source of opacity.

Pair production occurs at a threshold energy of $2m_e$ for the
photon in the rest frame of the nucleus, and gives an important
contribution only for the interaction with the CMB. We have
tabulated the pair production energy loss rates from Chodorowski
{\it et al.}~\cite{CZS} and treat them as continuous
losses~\cite{bs}.

The rate for photodisintegration is given by~\cite{PSB}:
\begin{equation}
  R_{A,i}=\frac{1}{2{\Gamma}^2}{\int}_0^{\infty}\,
  \,\frac{d\epsilon}{{\epsilon}^2}n(\epsilon){\int}_0^{2\Gamma\epsilon}
  \,\,\,d{\epsilon}^{\prime}{\epsilon}^{\prime}
  \sigma_{A,i}({\epsilon}^{\prime})\,,\label{EQ1}
\end{equation}
where $A$ is the atomic mass of the nucleus and $i$ is the number of
nucleons emitted. The Lorentz factor of the nucleus is given by
$\Gamma=E/(A m_p c^2)$ of the nucleus, $\epsilon$ and
$\epsilon^\prime$ are the background photon energy in the observer
frame and in the rest frame of the nucleus, respectively, and
$n(\epsilon)$ is the photon density of the ambient radiation.

The range of energies for the photodisintegration process, in terms of
the photon energy $\epsilon^\prime$ in the rest frame of the nucleus,
splits into two parts. The first contribution comes from the low
energy range up to 30 MeV, in the Giant Dipole Resonance region, where
emission of one or two nucleons dominates; the second contribution
comes from energies between 30 MeV and 150 MeV, where multi-nucleon
energy losses are involved. Above 150 MeV, following
\cite{PSB,salamon,Epele} we approximate the photodisintegration rates
by zero. This energy corresponds to the threshold for photopion
production, and we include this loss by using the cross-section of
nucleon photopion production scaled by the geometrical factor
$A^{2/3}$. Note that the energy carried away by a pion in such an
interaction is $\sim20$\% of the interaction nucleon energy, hence
only $\sim20$\%/$A$ of the primary nucleus. The threshold for
photopion production is also increased to $\simeq 4\,10^{19}\times
A$eV, and therefore pion production is only important for nuclei up to
$A\sim 4$.

Returning to photodisintegration, the lower limit of the integral on
$d{\epsilon}^{\prime}$, in Eq.~(\ref{EQ1}) was approximated by 2 MeV
for all reaction channels in PSB. We prefer to follow the approach of
Ref.~\cite{salamon}, with different thresholds for emission of one,
two and multiple nucleons, for different atomic numbers $A$. As
already discussed there, this could represent an important difference
compared to PSB because an increasing threshold energy may allow the
nucleus to propagate over longer distances.  For the cross sections
$\sigma_{A,i}({\epsilon}^{\prime})$ above threshold we used the same
parametrization as PSB.

We include the contributions from three different components of the
photon background: the first is given by the infra-red photons emitted
by galaxies and extends from $\simeq3.0\times10^{-3}\,$eV to
$\simeq0.33\,$eV.  We used the new estimates obtained from the
emissivity of the IRAS galaxies~\cite{Malkan}. The second one is the
CMB, extending from $\simeq2.0\times10^{-6}\,$eV to $\simeq4.0\times
10^{-3}\,$eV, and the third is the universal radio background (URB)
extending from $\simeq3.0\times10^{-9}\,$eV to
$3.0\times10^{-6}\,$eV. Due to Galactic contamination the latter can
not be measured directly below $\simeq1\,$MHz, however, we verified
that even the highest theoretical estimates from summing over the
contributions from normal and radio-galaxies~\cite{BP97} result in a
negligible contribution to photodisintegration at the energies of
interest ($\lesssim 10^{21}\,$eV).  Finally, we neglected the optical
background because, as can be seen in Fig.~1 in Ref.~\cite{Epele}, it
has no significant effect.

In a photodisintegration event the changes in energy, $\Delta E$, and
atomic number, $\Delta A$, are related by $\Delta E/E=\Delta A/A$.
Thus, the energy loss time due to photodisintegration is given by
$A/R_{{\rm eff},A}$, where
\begin{equation}
R_{{\rm eff},A}=\frac{dA}{dt}=\sum_i\,i R_{A,i}\,.
\end{equation}

\begin{figure}[ht]
\includegraphics[width=0.48\textwidth,clip=true]{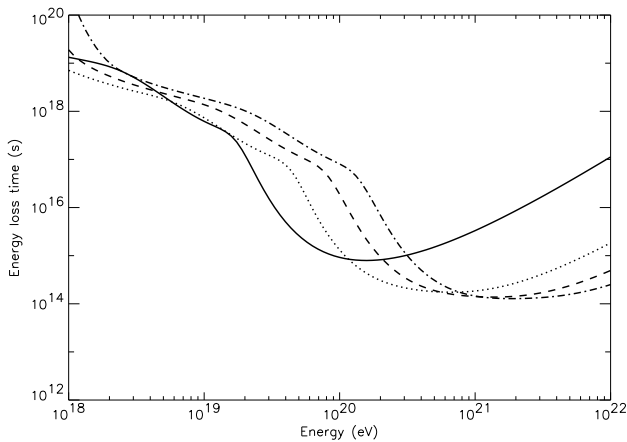}
\caption{The energy loss time vs energy for photodisintegration
on the combined CMB, infra-red and radio background.
The solid line is for Helium nuclei, the dotted line for Carbon,
the dashed line for Silicon and the dot-dashed line for Iron.}
\label{energyloss}
\end{figure}

In Fig.~\ref{energyloss} we show the energy loss time due to
single-nucleon, double-nucleon and multi-nucleon emissions in the
combined CMB, infra-red and radio background, for different atomic
numbers. We note that at energies above $10^{20}\,$eV the heaviest
nuclei start to disintegrate more quickly.  In addition, at these
energies, the multi-nucleon emission becomes more important compared
to one or two nucleon emission. Note also that the energy losses for
$^4$He (shown as a solid line in Fig.~\ref{energyloss}) do not include
photopion production, that become significant for energies $\gtrsim
1.5\times 10^{20}\,$eV.

\section{Numerical Simulations}
We use the same numerical approach used in earlier
publications~\cite{SLB99,LSB99,isola}, putting a single source at the
center, and we register all nuclei and nucleons arriving on 10
``detector'' spheres surrounding the source at radii scaled
logarithmically between 1.5 and 50 Mpc.  If not indicated otherwise,
we also assume that the source injects Iron nuclei with a $E^{-2}$
spectrum extending up to $\simeq10^{22}\,$eV.

We assume a homogeneous random turbulent magnetic field with power
spectrum $\langle B(k)^2\rangle\propto k^{n_B}$ for
$2\pi/L<k<2\pi/l_c$ and $\langle B^2(k)\rangle=0$ otherwise. We use
$n_B=-11/3$, corresponding to Kolmogorov turbulence, in which case
$L$, the largest eddy size, characterizes the coherence length of the
magnetic field. For the latter we use $L\simeq1\,$Mpc, corresponding
to about one turn-around in a Hubble time. Physically one expects
$l_c\ll L$, but numerical resolution limits us to
$l_c\gtrsim0.008L$. We use $l_c\simeq0.01\,$Mpc.  The magnetic field
modes are dialed on a grid in momentum space according to this
spectrum with random phases and are then Fourier transformed onto the
corresponding grid in location space. The r.m.s. strength $B$ is given
by $B^2=\int_0^\infty\,dk\,k^2\left\langle B^2(k)\right\rangle$.  The
simulations have been performed for two different strengths of the
magnetic field: a weak field corresponding to $10^{-12}\,$G and a
strong field corresponding to $2\times10^{-8}\,$G.

We injected $6\cdot 10^6$ Iron nuclei at the source. The equations
of motion in the presence of the magnetic force and the continuous
energy loss due to pair production are solved and at least every 0.01 Mpc
the nucleus is tested against photodisintegration
and photopion production, by using the rates determined as described
in the previous section.

We keep track of each individual secondary nucleus and each time such
a particle crosses one of the spheres of a given radius around the
source, arrival direction and energy are registered as one event on
this sphere. Energy loss processes and deflection are treated equally
for all produced secondary nuclei and nucleons.  In the diffusive
regime each trajectory is followed for a maximal time of 10 Gyr and is
abandoned if the particle reaches a linear distance from the source
that is twice the distance to the furthest sphere.


\section{Results}
We started our simulations injecting a distribution of Iron nuclei
following an $E^{-2}$ power law up to $10^{22}\,$eV.  We then followed
their disintegration history and kept track of all secondary nuclei
produced. We were thus able to evaluate the chemical composition of
detected nuclei at any given distance.

\begin{figure}[ht]
\includegraphics[width=0.48\textwidth,clip=true]{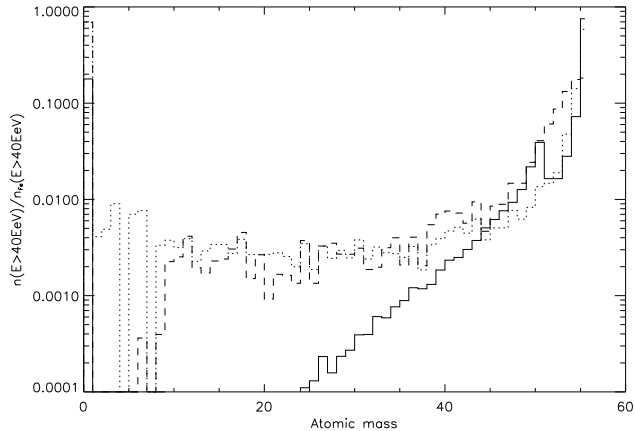}
\caption{Number of nuclei relative to Iron detected above
$4\times10^{19}\,$eV at three different distances
from the source as a function of their atomic mass for a field
$B=10^{-12}\,$G. Solid, dotted and dashed curves correspond to
distances $d=1.5\,$Mpc, 7.1 Mpc, and 50 Mpc, respectively.}
\label{istoW}
\end{figure}
\begin{figure}[ht]
\includegraphics[width=0.48\textwidth,clip=true]{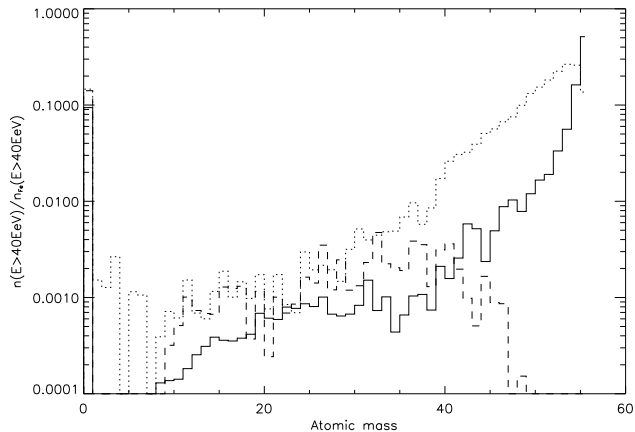}
\caption{Same as in Fig.~\ref{istoW} but for $B=2\times10^{-8}\,$G.}
\label{istoS}
\end{figure}

In Figs.~\ref{istoW} and~\ref{istoS} we show the chemical composition
of particles detected at three different distances for a magnetic
field of $10^{-12}\,$G and $2\times10^{-8}\,$G, respectively.  Results
are expressed as integral energy spectra $n(>E_{\rm th})$ of nuclei of
mass A detected above $E_{\rm th}$, as a function of A. We normalized
this quantity to the number $n_{\rm Fe}(>E_{\rm th})$ of (Iron) nuclei
emitted above the same energy.

In both figures one can see the increasing fraction of light nuclei
with increasing distance, and the progressive disintegration of Iron
nuclei. In the case of strong fields shown in Fig.~\ref{istoS}, heavy
nuclei are considerably deflected, which implies that the propagated
pathlength before reaching a given linear distance $d$ from the source
is much larger than $d$, the difference being more important for high
$Z$ and low $E$ nuclei (see spectra below). Diffusion increases the
number density of these nuclei due to their increased local residence
time, but it also increases their probability of photodisintegration
at a given distance. Therefore a strong magnetic field, {\it i.e.}
such that some UHECR enter a diffusion regime, not only modifies the
energy spectrum, it also modifies the chemical composition at a given
distance, with respect to the case of rectilinear propagation (small
deflection limit in a weak magnetic field).  Further effects of the
interplay between magnetic diffusion and energy losses will be shown
below.

\begin{figure}[ht]
\includegraphics[width=0.48\textwidth,clip=true]{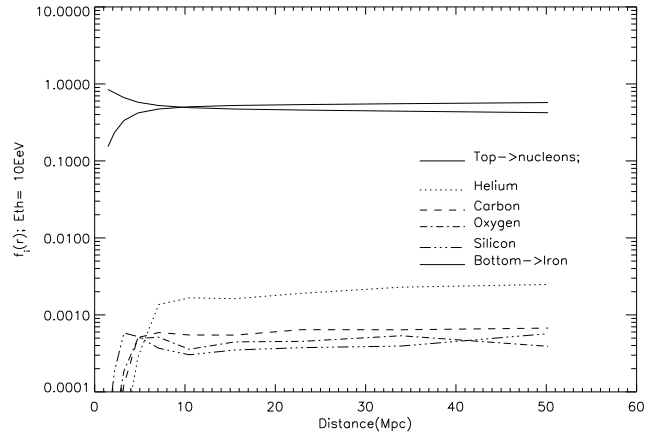}
\caption{Relative chemical composition above $10^{19}\,$eV
as a function of the distance for $B=10^{-12}\,$G.}
\label{chemW1}
\end{figure}

\begin{figure}[ht]
\includegraphics[width=0.48\textwidth,clip=true]{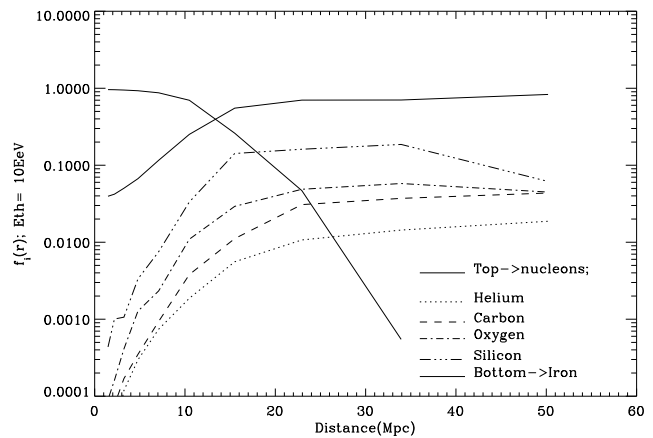}
\caption{Same as Fig.~\ref{chemW1} but for $B=2\times10^{-8}\,$G and
$E>10^{19}\,$eV.}
\label{chemS1}
\end{figure}

\begin{figure}[ht]
\includegraphics[width=0.48\textwidth,clip=true]{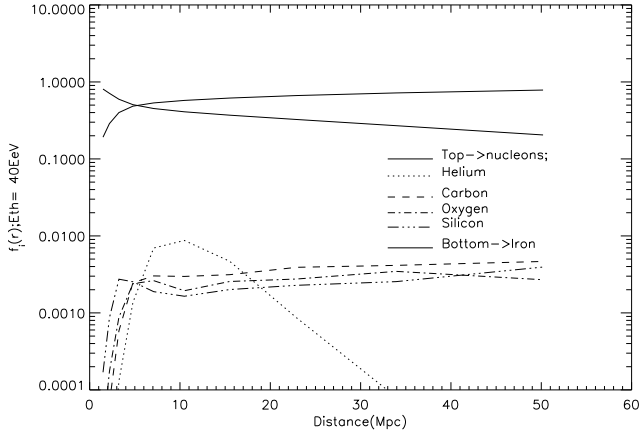}
\caption{Same as Fig.~\ref{chemW1} for a weak magnetic field but for
$E>4\times 10^{19}\,$eV.}
\label{chemW2}
\end{figure}

\begin{figure}[ht]
\includegraphics[width=0.48\textwidth,clip=true]{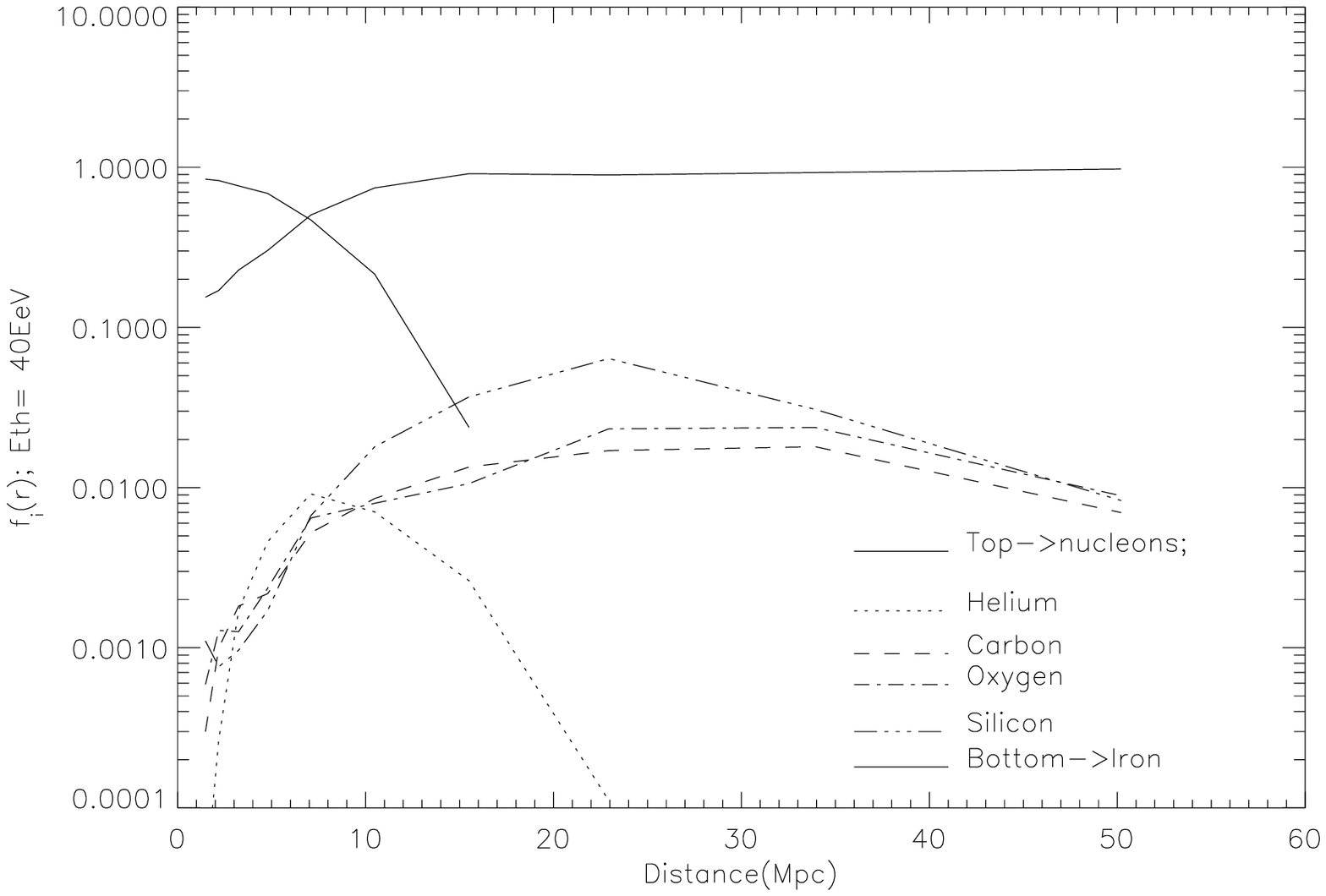}
\caption{Same as Fig.~\ref{chemS1} for a strong magnetic field but for
$E>4 \times 10^{19}\,$eV. A line for a given species that stops at
distance $<50\,$Mpc means that there is no particle left in that
species at greater distances in the simulation.}
\label{chemS2}
\end{figure}

To show the photodisintegration histories for different nuclei, we
plot in Figs.~\ref{chemW1} to~\ref{chemS2} the relative abundances
$f_i(d)$ of various atomic species as a function of distance $d$ for
different threshold energies, where $f_i(d)$ is defined as
\begin{equation}
  f_i(d)=\frac{n_i(d)}{\sum_i n_i(d)}\,,
\end{equation}
and $n_i(r)$ is the number of nuclei of species $i$ detected at
distance $d$ from the source.

\begin{figure}[ht]
\includegraphics[width=0.48\textwidth,clip=true]{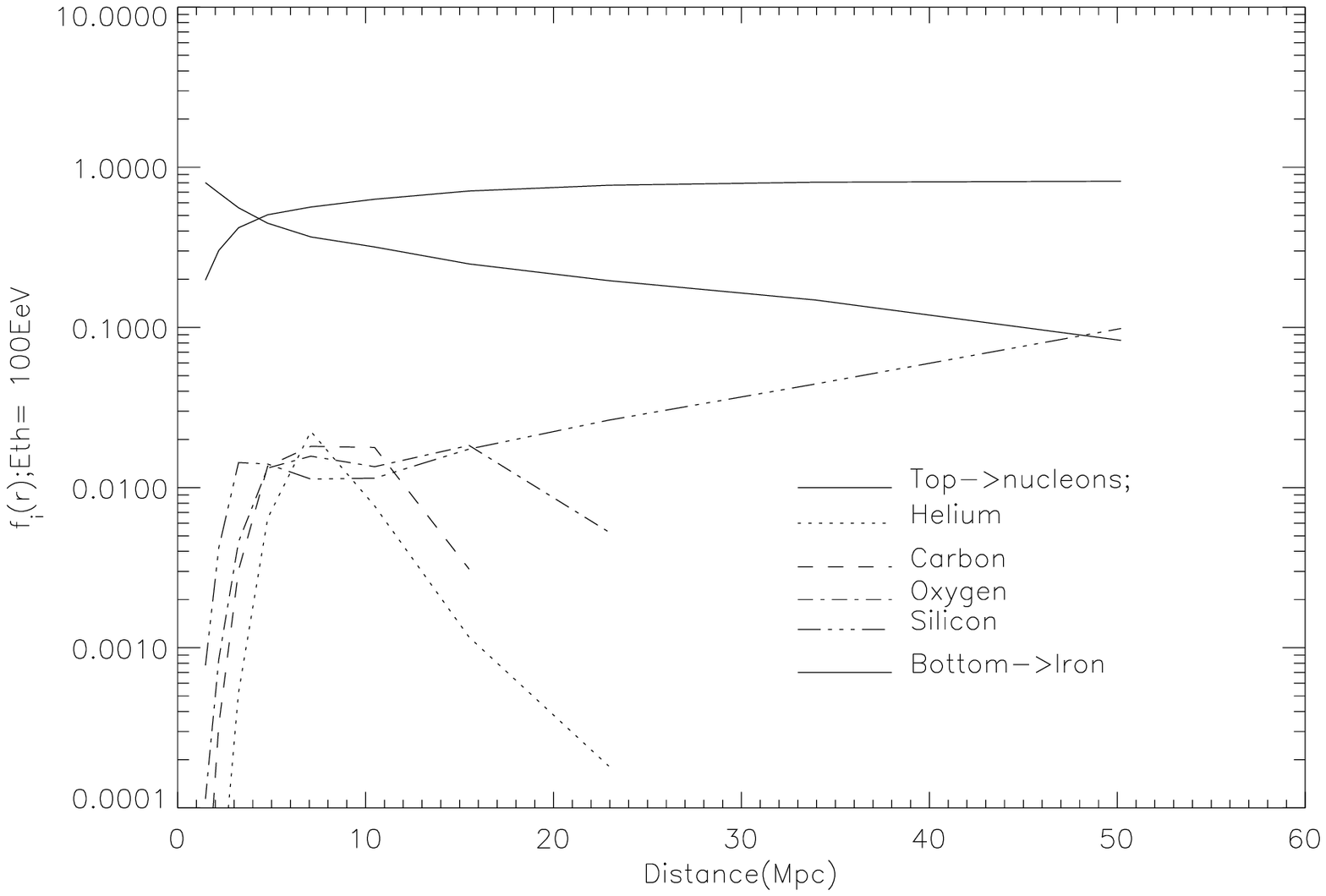}
\caption{Same as Fig.~\ref{chemW1} for a weak magnetic field but for
$E>10^{20}\,$eV.}
\label{chemW3}
\end{figure}

\begin{figure}[ht]
\includegraphics[width=0.48\textwidth,clip=true]{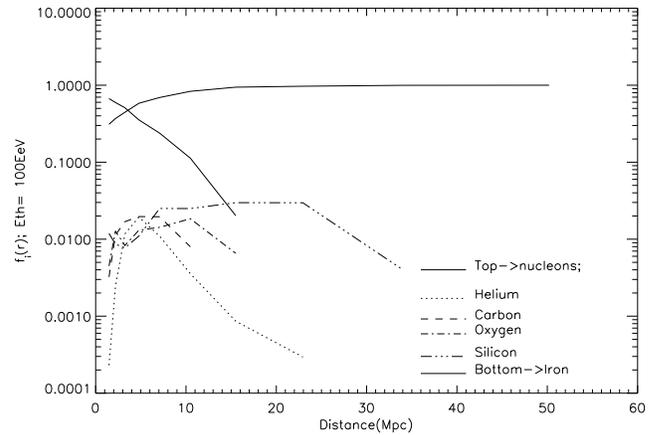}
\caption{Same as Fig.~\ref{chemS1} for a strong magnetic field, but
for $E>10^{20}\,$eV. A line for a given species that stops at distance
$<50\,$Mpc means that there is no particle left in that species at
greater distances in the simulation.}
\label{chemS3}
\end{figure}

As expected, Iron dominates the chemical composition at small
distances, whereas only protons are left for very large distances, in
agreement with previous studies on Iron nuclei propagation (in the
absence of a magnetic field). However Figs.~\ref{chemS3}
and~\ref{chemS2} show that UHECR above $10^{20}\,$eV cannot be
predominantly Iron at distances larger than
$\simeq10\,$Mpc~\cite{Epele} when propagating in a field of strength
$\simeq 2\cdot10^{-8}\,$G. In the case of a weak magnetic field, this
component can survive with a fraction $\gtrsim10$\% at all energies at
distances up to $\simeq50\,$Mpc. Again the effect of the magnetic
field is due to diffusion which increases significantly the effective
propagated distance for a given linear distance.

The effect of diffusion also becomes apparent by comparing the
relative abundances of heavy nuclei in Figs.~\ref{chemW1}
and~\ref{chemS1}: The case of a strong field shows an enhancement of
the relative abundances of Si, O and C by up to a factor 100 with
respect to the weak field case. At higher threshold energy this effect
becomes negligible because nuclei are no longer in the diffusive
regime. Light elements such as Helium are continuously produced by
photodisintegration of heavier nuclei, they reach a maximum relative
abundance, which we found to be around 1\%, then they quickly
disappear, reducing their abundance to 0.01 or 0.001\% at a distance
of 20 Mpc.

  One should note that the above figures are sensitive to the initial
maximum injection energy. In effect, here this energy $E_{\rm
max}=10^{22}\,$eV, which means that one cannot detect (secondary)
protons with energy $E>E_{\rm max}/56\simeq 1.8\cdot10^{20}\,$eV. If
the maximum injection energy is lowered, say $E_{\rm
max}\simeq10^{21}\,$eV, then one would not see protons with energy
$E\gtrsim 1.8\cdot10^{19}\,$eV, and consequently, in
Figs.~\ref{chemW2},\ref{chemW3},\ref{chemS2},\ref{chemS3} the chemical
composition would be dominated by Iron nuclei at all energies. In
Figs.~\ref{chemW1},\ref{chemS1} with threshold $E_{\rm
th}=10^{19}\,$eV, the proton domination would be reduced. Furthermore,
in Figs.\ref{chemS2},\ref{chemS3}, the composition would becomes
dominated by intermediate mass nuclei at distances $\gtrsim 15\,$Mpc.

  These figures also depend on the energy spectrum index
chosen. Indeed, it is easy to see that if the energy spectrum of
primary nuclei $N_{\rm p}(>E)\propto E^{1-\alpha}$, then if all nuclei
above energy $E$ are photodisintegrated in secondary protons of energy
$E/A$ (and above), the number ratio of secondaries $N_{\rm s}(>E/A)$
to primaries at the same energy $N_{\rm p}(>E/A)$ reads $N_{\rm
s}/N_{\rm p}=A^{2-\alpha}$ (all other losses neglected). Therefore,
depending on the spectral index $\alpha$ (taken here as $\alpha=2$),
the secondary flux has more or less importance compared to the primary
flux. For hard spectra $\alpha < 2$, the secondaries tend to dominate,
while the reverse is true for $\alpha >2$.

 This latter statement is obviously modified in the presence of a
strong magnetic field, since particles of a same energy but different
mass have a difference magnetic rigidity. As a consequence, at a same
energy, high $Z$ particles (in our case, primaries) may be diffusing
and their local density increased while low $Z$ particles (e.g., here
secondary protons) may be non-diffusing and their local density not
increased. In a strong magnetic field, for a hard injection spectrum
$\alpha<2$, one may thus see different regimes, in which either the
secondaries dominate (low energy, where both protons and Iron nuclei
diffuse, or high energy, where both protons and Iron nuclei do not
diffuse), or the primaries dominate (when protons do not diffuse but
Iron nuclei of the same energy diffuse). If $\alpha>2$, then
secondaries give a subdominant contribution in all cases.

\begin{figure}
\includegraphics[width=0.48\textwidth,clip=true]{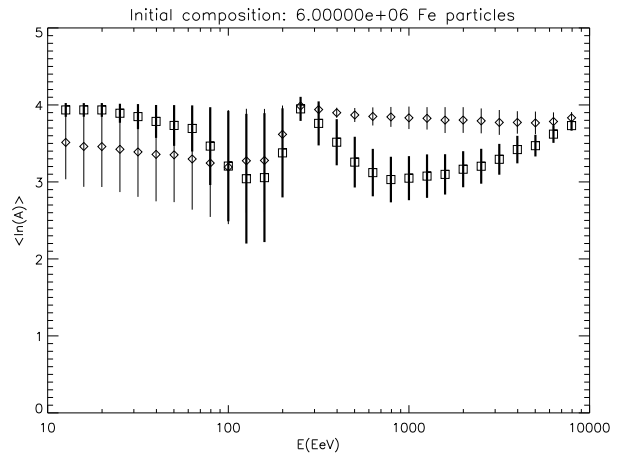}
\caption{Average logarithmic nucleus mass $A$ as a function of energy
for $B=10^{-12}\,$G (diamonds) and $B=2\times10^{-8}\,$G (squares)
at a distance $d=1.5\,$Mpc.}
\label{Amean1}
\end{figure}

\begin{figure}
\includegraphics[width=0.48\textwidth,clip=true]{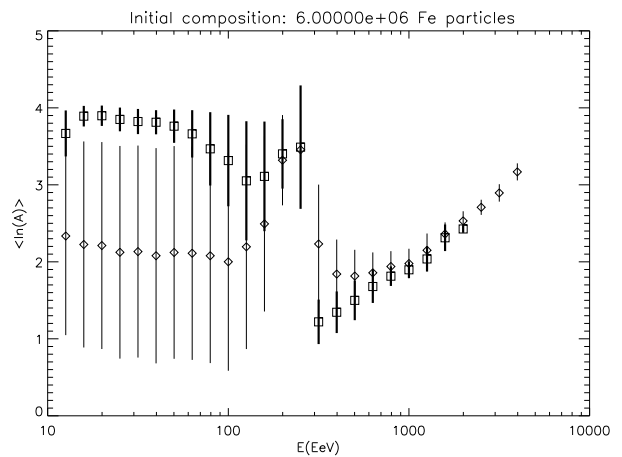}
\caption{Same as Fig.~\ref{Amean1} but at for $d=7.1\,$Mpc.}
\label{Amean2}
\end{figure}

To further investigate the diffusion problem and photodisintegration
processes we studied the energy dependence of the average mass and the
observed spectra at different distances from the source.
Figs.~\ref{Amean1} and~\ref{Amean2} show the average detected
logarithmic nucleus mass $\log A$, as a function of energy, for two
different distances from the source. The sudden change of the plots at
an energy around $2\times 10^{20}\,$eV is also due to the maximum
injection energy which translates here for a maximum proton energy
$\simeq1.8\times10^{20}\,$eV.

In these figures, we see that at low energies $\lesssim10^{20}\,$eV
the average composition is more strongly dominated by Iron nuclei in
the strong magnetic field case than in the weak magnetic field
case. This is an effect of diffusion, as before, which increases the
local density of diffusing particles {\it vs} that of non-diffusing
particles. While Iron nuclei of energy $\lesssim 10^{20}\,$eV diffuse
in $B\simeq2\cdot10^{-8}\,$G, protons of the same energy do not
diffuse, hence the effective enhancement of Iron nuclei with respect
to secondary protons. Here as well, note that the above conclusion
depends on the spectral index chosen. If the spectrum is hard
($\alpha<2$) then the importance of the secondary proton is increased
with respect to that of the primary nuclei flux, and the above effect
is reduced.

At higher energies, an opposite effect happens, {\it i.e.}, the
composition is lighter for a stronger field, because
photodisintegration is more important than at low energies and
increases with the larger propagated pathlength in stronger fields.
For larger distances (see Fig.~\ref{Amean2}) a continuous increase of
the average logarithm of $A$ is seen above the high energy proton
cut-off for both field strengths. This is due to the fact that the
maximum energy that one can detect for a species of mass $A$ increases
with $A$, as discussed above, which implies that moving toward higher
energies we select heavier nuclei.

 Note that here as well these figures depend rather strongly on the
initial maximum injection energy. Moreover, the rather large error
bars on the average logarithmic mass are not really representative of
a gaussian standard deviation, since the mass distribution is strongly
peaked on Iron nuclei and protons. The large error bar simply reflects
the large mass difference these two peaks. This two-peak behavior can
be seen in Figs.~\ref{istoW},\ref{istoS} which show the distribution
in mass of the composition above $\cdot10^{19}\,$eV for various
distances: one clearly see in these figures that most recorded
particles are either protons or iron nuclei. This effect is reduced in
the case of a strong magnetic field (as photodisintegration losses are
more severe due to increased effective length traveled), as shown in
these figures and by the reduced size of error bars on the average
$\log A$ in Fig.~\ref{Amean1},\ref{Amean2}.

\begin{figure}[ht]
\includegraphics[width=0.48\textwidth,clip=true]{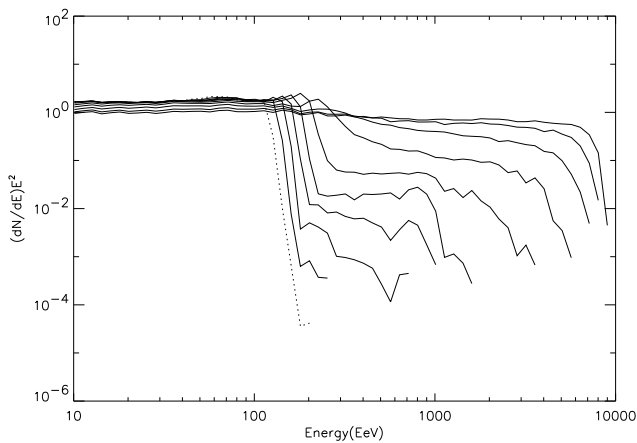}
\caption{All-particle spectrum observed at distances $d=1.5, 2.3, 3.2,
4.8, 7.1, 10.5, 15.5, 23., 33.9, 50$ Mpc from right to left. The
dotted line is for $d=50\,$Mpc, and $B=10^{-12}\,$G.}
\label{speW}
\end{figure}

\begin{figure}[ht]
\includegraphics[width=0.48\textwidth,clip=true]{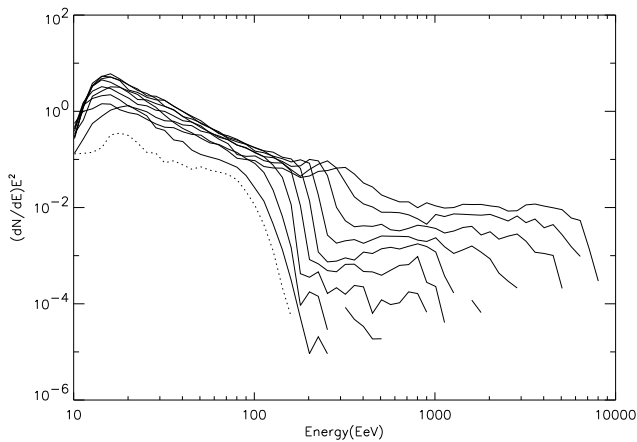}
\caption{Same as Fig.~\ref{speW} but for $B=2\times10^{-8}\,$G.}
\label{speS}
\end{figure}

We finally show in Figs.~\ref{speW} and~\ref{speS} the expected
spectra at different distances for the two field strengths.
Fig.~\ref{speW} shows the weak field case and can be compared to
Fig.~3 of Ref.~\cite{salamon} (although these authors chose to use a
spectral index $\alpha=3$). Figure~\ref{speW} shows a cut-off at
energy $E\simeq1.5-2\cdot10^{20}\,$eV that is increasingly pronounced
with distance, in agreement with previous
works~\cite{PSB,salamon,Epele}.  Figure~\ref{speS} shows a
characteristic spectral slope due to diffusion of nuclei in the
magnetic field for energies below the cut-off, and an almost flat
component at highest energies (recovery of the injection spectrum in
absence of losses, for rectilinear propagation).  The fact that the
transition energy between diffusive and rectilinear propagation occurs
around the cut-off energy $\sim10^{20}\,$eV is due to the choice of
the magnetic field strength $B\simeq20\,$nG. If the magnetic field
were substantially stronger, the increased length traveled for
particles above the cut-off would result in a more pronounced cut-off
for a same distance.

In Fig.~\ref{speS} one also notes the presence of a low energy
cut-off around $E\sim1.5\cdot10^{19}\,$eV. This is due to the fact
that the Larmor radius of Iron nuclei at energies around $10^{19}$
eV and in a magnetic field around $2 \times 10^{-8}$G is about
$2\cdot10^4$pc, comparable with $l_c$, which represents the
resolution of our numerical simulation. Furthermore, we follow
nuclei up to a maximum propagation time of the order of the age of
the Universe, $\tau_{max} = 10^{10}$ Gyrs; particles with high Z
and low energies can have a propagation time larger than
$\tau_{max}$ and never reach distant shells. This effect
represents an additional contribution to the low-energy cut-off of
spectra on distant shells.


\section{Applications}
Let us now apply the results of our numerical simulations to test some
models recently proposed to explain the origin of cosmic rays.

\subsection{Iron nuclei from nearby starburst galaxies}

In a recent work Anchordoqui et al.~\cite{anchor} have put forward
the possibility that cosmic rays above the ankle are essentially
heavy nuclei which originate in two nearby ($d\sim3\,$Mpc)
sources, the starburst galaxies M82 and NGC 253, and propagate in
a $B\simeq15\,$nG extra-galactic magnetic field which isotropize
the arrival directions on Earth. They based their analysis on
analytical estimates of the diffusion coefficient and
approximations to the photodisintegration losses and angular
deflections. Our numerical simulations are well suited to improve
the discussion of their hypothesis, thanks to a more accurate
treatment of photodisintegration processes and to a treatment of
deflection without approximations.  One should first note that
Fig.~\ref{Amean1} shows that starting with a distribution of Iron
nuclei at a linear distance $d\simeq3\,$Mpc with $B\simeq20\,$nG,
the average nucleus mass is still high: $\log A\simeq 3-4$ at
$E\simeq 10^{20}\,$eV. This suggests that the heavy component can
survive across this distance; this is in agreement with the
results of Ref.~\cite{anchor}.

\begin{figure}[ht]
\includegraphics[width=0.48\textwidth,clip=true]{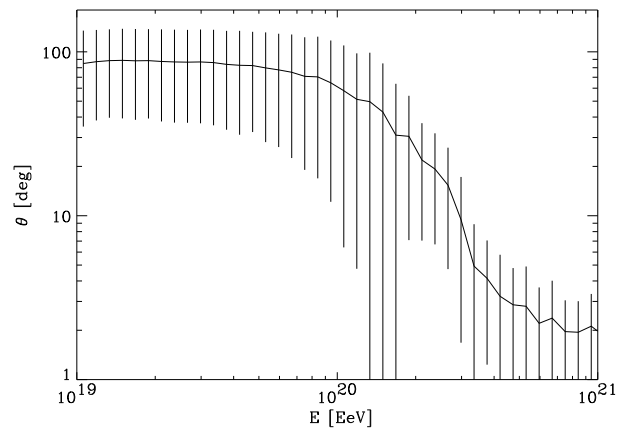}
\caption{The average angular deflexion vs. energy for a source at a
distance $d=3.2\,$Mpc.}
\label{martinplot}
\end{figure}

In Fig.~\ref{martinplot} we show the angular deflection, defined
as the angle between the source direction and the momentum of the
particle when it is recorded, as a function of energy. Here as
well, the sudden increase of error bars around
$2\times10^{20}\,$eV is due to the presence of secondary protons
in the signal; protons with $E\sim10^{20}\,$eV suffer a similar
deflection than iron nuclei of energy $\sim 2\times 10^{21}\,$eV,
which is of order a few degrees. The same angular deflection when
plotted {\it vs} magnetic rigidity $R\equiv E/Z$ shows a much more
regular behavior, similar to that shown in Fig.~\ref{martinplot}
up to the size of the error bars.

This figure shows furthermore that for energies below the transition
energy, the arrival directions have been isotropized as $\theta \sim
90^\circ \pm 90^\circ$. In the high energy regime, one recovers a
power law behavior $\theta \propto E^{-1}$ but we find an average
angular deflection that is overall a factor $\simeq4$ from that given
analytically from a random walk argument by Waxman \&
Miralda-Escud\'e~\cite{WM96}, and used by Anchordoqui {\it et al.}
~\cite{anchor}. We believe this difference is due to order of unity
factors entering the random walk formula and our convention of
defining the coherence length, and the range of applicability of the
random walk formula.


Of interest is the prediction of an anisotropy that should be seen at
the highest energies, $E\sim 2-3\cdot 10^{20}\,$eV as suggested by
Fig.~\ref{martinplot}. We also note that the highest energy Fly's Eye
event Fly's Eye event of energy $E=3.2\pm 0.9\times10^{20}$ eV arrived
from a direction that is $\simeq37^{\circ}$ away from M82 (see
Anchordoqui et al.~\cite{anchor}, and references therein). By
comparing with Fig.~\ref{martinplot}, this event appears to be only in
marginal agreement with our simulation, but the rather large
deflection could be explained by a slight overestimate of the
energy. Note also that the Fly's Eye event is located $98^\circ$ away
from NGC~253.

Two other very high energy events have been reported by the AGASA
experiment, one with $E\simeq2.1\pm 0.6\cdot10^{20}\,$eV with
arrival direction
$(\alpha,\delta)=(19^\circ,+21^\circ)$~\cite{AGASA} (equatorial
coordinates), the other with $E\sim 3\times10^{20}\,$eV and
arrival direction $(\alpha,\delta)=(359^\circ,22^\circ)$ (this
latter is preliminary, see Ref.~\cite{AGASA_ICRC}). These two
events are located at $82^\circ$ from M82 and $47^\circ$ from
NGC~253 for the former, and $86^\circ$ from M82 and $49^\circ$
from NGC~253 for the latter. For these two events as well the
agreement with Fig.~\ref{martinplot} is marginal, although
slightly better respect to the Fly's Eye event.


\begin{figure}[ht]
\includegraphics[width=0.48\textwidth,clip=true]{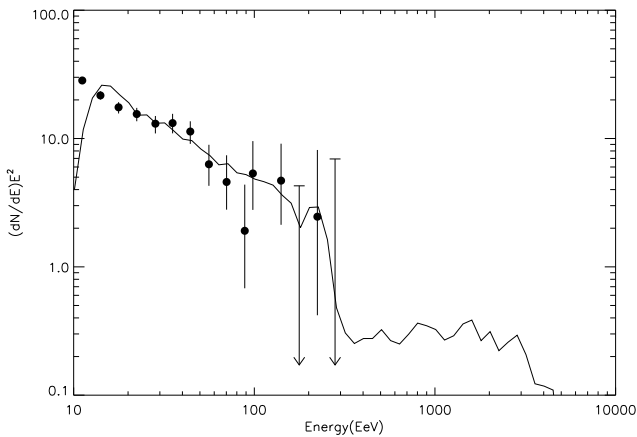}
\caption{Observed spectrum at 3.2 Mpc for B=$2 \times 10^{-8}$G,
compared with AGASA data. Injection spectrum $\propto E^{-1.6}$
and normalization was fit to the data.}
\label{spagasa}
\end{figure}

Finally, in Fig.~\ref{spagasa} we compare the spectrum observed at
distance $d=3.2\,$Mpc from a single source for a magnetic field
$B=20\,$nG with the observed AGASA spectrum. It turns out that in
order to fit the AGASA spectrum an injection spectrum $\propto
E^{-1.6}$ is required. This is relatively hard compared to the
$E^{-2}$ injection spectrum usually expected for shock
acceleration~\cite{biermann}. The hardness of the spectrum required is
likely due to the interplay between energy losses, existence of
secondaries and diffusion at low energies.

Note that it is not realistic a priori to expect that a source
such as a starburst galaxy would accelerate only iron nuclei, and
not lighter nuclei. In particular the cosmic abundance of iron
would suggest that protons should be much more abundant at the
same rigidity. Assuming that the accelerated spectrum for species
$i$ as a function of rigidity $R$ can be written ${\rm d}n_i/{\rm
d}R = N_i (R/R_0)^{-\alpha}$, one finds that the ratio of fluxes
of two species $i$, $j$ at a given energy reads $F_i/F_j =
(Z_i/Z_j)^{\alpha-1} N_i/N_j$. If $i$ corresponds to protons, $j$
to iron nuclei, and $\alpha=1.6$, then $F_p/F_{\rm Fe}\simeq 0.1
N_{\rm p}/N_{\rm Fe}$. If $N_{\rm p}/N_{\rm Fe}$ is corresponds to
the cosmic abundance of iron, then indeed one cannot consider iron
as the dominant species. However in the present scenario, it is
assumed that acceleration takes place in two steps, first in
supernovae shock waves up to $10^{15}\,$eV then reaccelerated in
the galactic wind up to $10^{20}\,$eV. It is not clear in this
case to what $N_{\rm p}/N_{\rm Fe}$ refers, but if in a first
approximation one considers that it is the proton to iron ratio at
$\sim10^{15}\,$eV, this latter is found to be of order a few for
galactic cosmic rays. This is mainly due to the fact that the
spectrum of heavier nuclei cosmic rays is generically harder than
that of lighter nuclei. In that case, indeed the contribution of
protons, and for that matter, or intermediate mass nuclei, to the
energy spectrum at injection can probably be neglected in a first
approximation.

Finally, one should note that starburst galaxies are active for a
finite amount of time: only $\sim10^8\,$yrs. Here for
$B\simeq20\,$nG the time delay at $E\sim 10^{19}\,$eV is of order
of a few $10^8\,$yrs. If the high energy part of the spectrum has
been recorded (in part) by the present experiment, then the flux
at the lower end of the spectrum should be depleted in heavy
nuclei, as most of these particles would not have had enough time
to reach us. It is difficult to quantify this effect at present,
but it constitutes a potential signature of this scenario for
future detectors.

\subsection{Helium nuclei from M87}

In a different scenario, proposed by Ahn et al.~\cite{Ahn}, M87 in the
Virgo cluster (located at a distance $d\approx20\,$Mpc from the Milky
Way), is assumed to be the local source of UHECR. Indeed the authors
showed that one can trace back to M87 the 13 events observed above
$10^{20}$ eV~\cite{AGASA} if the Galactic magnetic field has the
structure of a Parker spiral and extends to $\sim 1-2\,$Mpc. They
showed that provided the two highest energy events are Helium nuclei
and the others protons, all 13 events point back to within
$20^{\circ}$ of M87. The importance of the specific magnetic field
chosen to reach this conclusion was stressed in a note by Billoir and
Letessier-Selvon~\cite{billoir}.

\begin{figure}[ht]
\includegraphics[width=0.48\textwidth,clip=true]{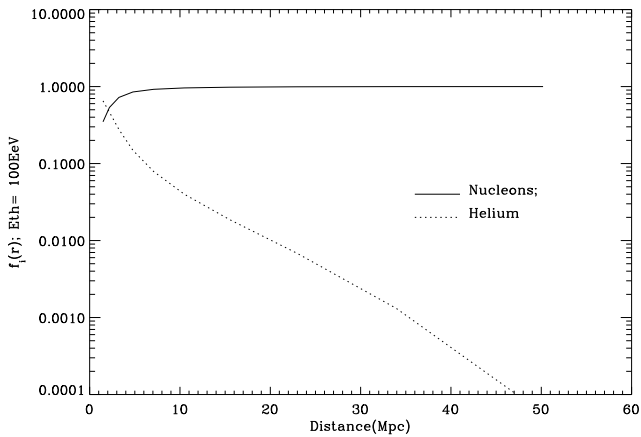}
\caption{Relative chemical composition as a function of the distance
for a simulation with a source injecting Helium nuclei, for $B=10^{-12}\,$G.}
\label{chemhelium}
\end{figure}

To see if such a composition is possible we studied the relative
abundance of Helium as a function of the distance from the source. In
Fig.~\ref{chemhelium} we show the chemical composition as a function
of distance, assuming that only He nuclei are injected in M87 and that
the extra-galactic magnetic field $B=10^{-12}\,$G.

This shows that even for very weak fields and thus negligible
deflection, the abundance of Helium nuclei with energy above
$10^{20}\,$eV is a factor 100 smaller than the nucleon abundance at
distances $d\simeq20\,$Mpc.  The probability of observing two helium
nuclei out of 11 protons in this energy range is thus extremely small.
If all the events are protons, the convergence in the direction of M87
is poor, which makes the model much less attractive. To start with a
distribution of Iron nuclei would make things worse, as can be seen in
Fig.~\ref{chemW2}, leading to Helium nuclei abundances at least $10^3$
times smaller than the nucleon abundance.

Although we considered a weak stochastic magnetic field while Ahn {\it
et al.} considered a strong coherent magnetic field structured as a
Parker spiral, this should not make a big difference in our
conclusions. As a matter of fact, if one increases the magnetic field
strength, the effective length traveled for He nuclei is increased,
hence photodisintegration should be more severe. This scenario thus
appears fine-tuned in so far as the chemical composition is concerned.

\section{Conclusions}
In this paper we studied the propagation of a distribution of
heavy nuclei in a stochastic magnetic field, including all
relevant energy loss processes.  For the propagation in the
magnetic field we used the same numerical approach as in
Ref.~\cite{SLB99,LSB99,isola}. This approach was here generalized
to heavy nuclei and their photodisintegration processes.  One main
conclusion of this paper is that a strong magnetic field, i.e.
such that some UHECRs experience a diffusive propagation regime,
can strongly modify the chemical composition and energy spectrum
at a given energy with respect to what would be seen in the
absence of a magnetic field. Rather generically, an increased
magnetic field implies a larger effective length travelled, hence
a larger photodisintegration probability, hence a chemical
composition shifted to ligther species. As we have argued, the
extent of this effect also depends on the injection spectrum
spectral index, and on the maximal injection energy. If the
injection spectrum ${\rm d}n/{\rm d}E \propto E^{-\alpha}$, then
if $\alpha>2$ the secondary protons produced in
photodisintegration interactions do not give a dominant
contribution in the low energy observed flux. The converse is not
generally true in the case of a strong magnetic field, as the
injection spectrum is softened by diffusion.


We applied our results to the discussion of two models recently
proposed to explain the origin of UHECR. Our simulations suggest that
the model proposed by Anchordoqui et al.~\cite{anchor}, in which UHECR
are iron nuclei accelerated in nearby starburst galaxies, is in
relatively good agreement with the data as far as the energy spectrum
is concerned. However, it requires a relatively hard injection
spectrum ($\alpha\simeq1.6$), and the three highest energy events from
AGASA and Fly's Eye are $\gtrsim40^\circ$ away from the galaxies
proposed as sources, in marginal agreement with the expected
deflection.

We also showed that for an injection spectrum dominated by Helium
nuclei, the relative abundance of Helium compared to nucleons turns
out to be smaller than $0.01$ at distances $\sim20\,$Mpc from the
source. This implies that the scenario of Ahn {\it et al.}~\cite{Ahn},
which suggests that the UHECR originate from M87 and are deflected in
a powerful Parker spiral Galactic magnetic field, and which requires
that the two highest energy cosmic rays (out of 13 above
$10^{20}\,$eV) are He nuclei, is highly fine-tuned.

\end{document}